# STAKEHOLDER TOOL FOR ASSESSING RADIOACTIVE TRANSPORTATION (START) VERIFICATION AND VALIDATION EFFORTS

Caitlin Condon, Kacey McGee, Harish Gadey, Patrick Royer

Pacific Northwest National Laboratory, Richland, Washington, United States of America 99354

**ABSTRACT**

The U.S. Department of Energy (DOE) Office of Integrated Waste Management is planning for the eventual transportation, storage, and disposal of spent nuclear fuel (SNF) and high-level radioactive waste (HLW) from nuclear power plants and DOE sites. The Stakeholder Tool for Assessing Radioactive Transportation (START) is a web-based, geospatial decision-support tool developed for evaluating routing options and other aspects of transporting SNF and HLW, covering rail, truck, barge, and intermodal infrastructure and operations in the continental United States. The verification and validation (V&V) process is intended to independently assess START and provide confidence in the ability of START to accurately provide intended results. The results selected for independent V&V of the START code include those identified as crucial outputs of the START code by subject matter experts.

The V&V efforts included an independent analysis of select outputs through other geographic information system (GIS) programs including QGIS. QGIS is an open-source GIS tool that is independent of the ESRI GIS server used within START, thereby allowing for an independent V&V analysis. The V&V efforts have focused primarily on the route buffer zone population values as well as the route lengths reported in START. The efforts include developing a V&V methodology for each output type that an independent user could replicate through the user interface for a small number of routes as well as more automated methodologies that reduce the number of interactions with the GIS user interfaces. Independent analysis of the routes showed excellent agreement between the START outputs for population within the route buffer zone and the length of the routes. Over 200 test routes were evaluated; in all cases, the percent difference in the population within the buffer region was less than +/- 5%, and most cases were below +/- 1%. The V&V work is ongoing and other START features that are undergoing independent analysis include population density within the route buffer zone as well as reported doses for the segments within the START code. PNNL-SA-184470.

**INTRODUCTION**

Within the United States, much of the spent nuclear fuel (SNF) is currently stored on site at its respective point of origin. The Department of Energy (DOE) continues to plan for the eventual transportation, storage, and disposal of SNF and high-level radioactive waste (HLW) from nuclear power plants and DOE sites [1]. To support this effort, the DOE continues the development of the Stakeholder Tool for Assessing Radioactive Transportation (START) software as an assessment and decision-making tool [2].

START supports many aspects of the DOE Integrate Waste Management (IWM) program, including serving as a communications tool for conveying geospatial data and information, an options analysis tool for exploring potential transport modes and routes, an emergency response planning tool for States and Tribes to identify training needs along potential transport corridors, an environmental analysis tool for estimating potential radiation dose exposure from incident-free and incident-case transport conditions, and a systems analysis support tool for providing route-related inputs for throughput analysis.

The DOE anticipates that START will have a wide user base that will include Federal, State, Tribal, and local government officials as well as nuclear utilities, transportation carriers, support contractors, and other stakeholders. For this reason, START has been designed to enable representation of a wide range of operating scenarios and performance objectives, with an emphasis on allowing flexibility. START includes a wide variety of GIS Data Layers including layers with information such as potential shipment origin and transfer points, educational and elderly care facilities, environmental land uses, emergency response assets, transportation infrastructure and operations, political jurisdictions, mass gathering places, and existing routes.

START is a web-based tool using a Geographic Information System (GIS) server to generate possible transportation routes. Figure 1 shows an example of the START user interface before a user has generated a route. When generating the route, users are able to select an origin, destination, and transportation method along with any additional stops or route avoidances. Route criteria can be selected as well, forcing certain aspects of the route to take priority. This option includes generating a path that minimizes population along the route or path that minimizes the distance. These features enable the user to customize and easily generate multiple routes for comparison.

START focuses on moving commercial SNF from nuclear power plants and HLW stored at DOE sites to alternative destinations. At present, the primary mode for these shipments is expected to be via rail. For many shipment sites, access to the national rail network will typically require initial use of a local/regional (short line) railroad or involve intermodal shipments where the access leg is a movement performed by heavy-haul truck and/or by barge.

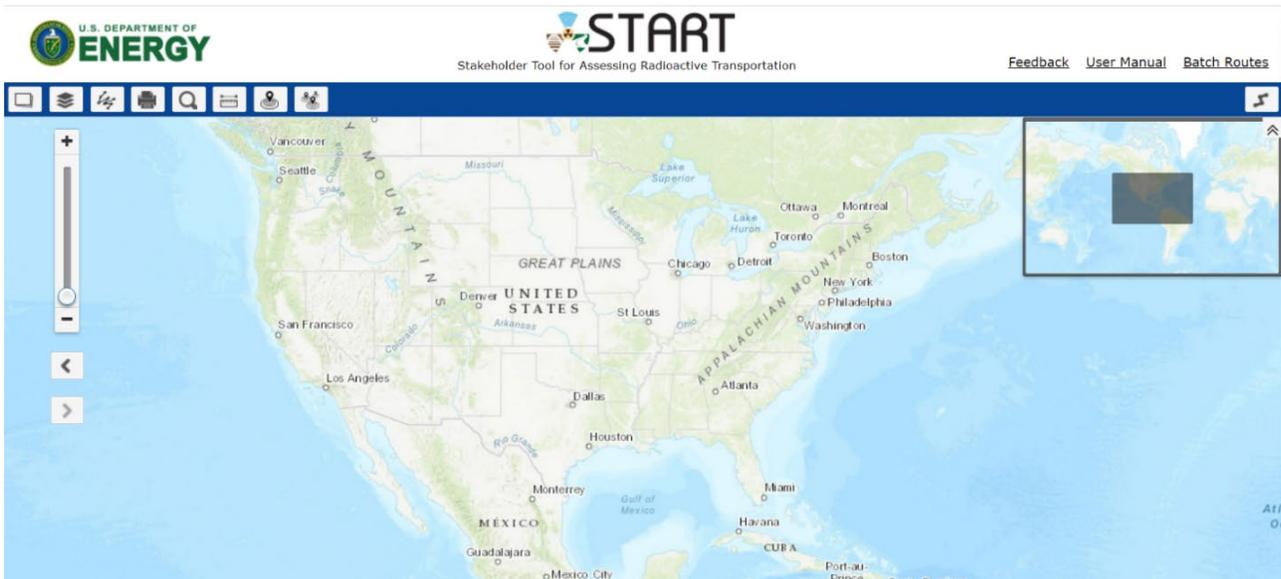

**Figure 1. Stakeholder Tool for Assessing Radioactive Transportation (START) interface**

Once generated in START, routes provide a detailed summary for the overall route as well as for the individual route segments. Overall route details allow for the direct comparison between routes based on information such as route length, population along the route, and cumulative accident probability. The segment details allow for a finer adjustment to be made to a route such as avoiding specific areas or population density centers. These details and features are designed to provide users with a flexible tool that can aid in the planning and decision-making process.

Routes generated in START also produce output files available to the user for download. These files give the option for further analysis within other software programs. Each route generates a shapefile, KML file, as well as multiple CSV files as output. The shapefile and KML files allow for further analysis using other GIS programs such as ArcMap, Google Earth or QGIS [3]. Among the CSV files are the summary details file and the route details files. The summary details CSV allows for direct comparisons of overall route statistics while the route details CSVs allow for evaluation of the route segments and allow for finer adjustments to be made to the route.

Throughout the development process, START is updated using an agile style of software development with collaboration between the development team and an independent V&V team. Working together, the development team provides new and updated features implemented in each new version of START that are subsequently evaluated by the V&V Team. The V&V team then provides feedback and suggestions for further updates for future versions. This cycle enables independent analysis of features and feedback, providing a more robust and reliable software package with each iteration. This cycle is depicted in Figure 2, the conceptual model of the verification and validation process.

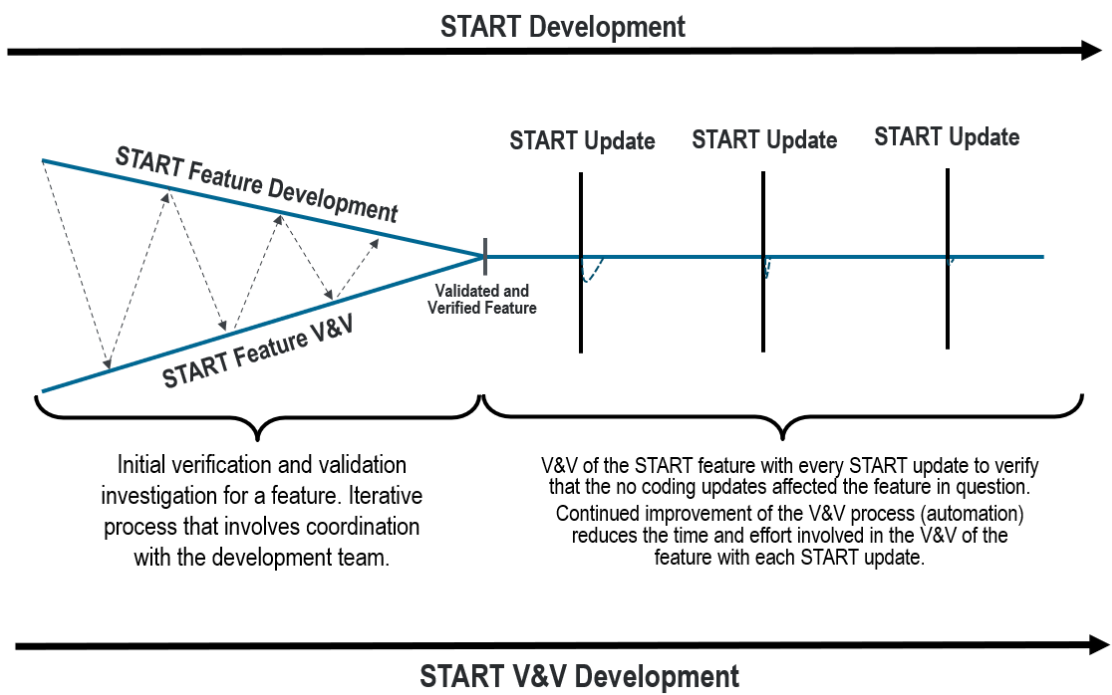

Figure 2. Conceptual figure of the verification and validation process

An ESRI GIS server is used to generate routes within START. To avoid bias when verifying data from start, the V&V team uses an unaffiliated GIS software, QGIS, for analysis [4]. Route data is generated within START and then, through an output file, imported into QGIS for analysis. This was done to ensure the data provided by START is in good agreement with data provided by independent software.

**DESCRIPTION**

The independent V&V of START began in 2020. During that first year, the V&V team began to evaluate crucial START outputs and developed the V&V conceptual model show in Figure 2. V&V of START is ongoing as START continues to be developed and new data features are added or updated. As START implements updates, output data that is considered crucial is validated against the previous version's data as well as any updated data layers and new features are evaluated for accuracy. V&V analysis of critical data outputs have been conducted for START versions 3.2.1, 3.2.2., and START version 3.3.

This paper covers the V&V analysis for START version 3.2.2. A separate paper will cover updates of START V&V activities for version 3.3 to date. With the update of START v. 3.2.2, the V&V focused on verifying length and population within the buffer; other outputs were not included in this V&V analysis.

For data validation, a series of test routes were used. Routes originated from DOE, or commercial nuclear Facilities within the US and had a hypothetical destination of Lebanon, KS, chosen as it is the geographic center of the contiguous US (GCUS). Test routes included 155 unique origin/ destination pairs, each evaluated with 800m and 2500m buffer zones, for a total of 310 routes. Once generated, the test route output files were imported into QGIS for analysis. These test routes have been used in previous versions of START, allowing for direct comparisons to be made between different START versions.

**START POPULATION DATA**

Population within the buffer zone was identified as a crucial output of START that could be used in an impact evaluation and so is included in START V&V analysis. When generating a route, START calculates population data using a designated buffer zone along the route and determines the geographic overlay with a population raster data set. Using the same method, test routes were imported into QGIS where a buffer zone was created and overlayed with a population raster data set. The route population data for START v.3.2.2 was compared to an independent analysis using the same data sets and recreating the workflow in QGIS. Workflows were developed and Python scripts were written to partially automate the process for the V&V efforts. Population data was extracted for all the routes for both 800-meter and 2500-meter buffer zone distances from the LandScan raster population data sets [4] for both day and at night. The average population for a route was calculated using Eq. 1.

$$\text{Average Population} = \frac{Day + (2*Night)}{3} \qquad \text{(Eq. 1)}$$

The average population extracted from QGIS and ArcMap were then compared to the START results as part of the V&V process. Strong agreement was observed between the START results and the ArcMap data; this result was expected as START uses ESRI features for calculations and ArcMap is an ESRI product. The results can be seen in Figure 3 and Figure 4, START population data comparison to calculated population data using LandScan data inputs. The results showed good overall agreement with an average difference of 0.24% (SD 0.29%). The data was then broken into subsets based on transportation method to determine if any subsets showed greater differences. All subsets were in line with the overall differences.

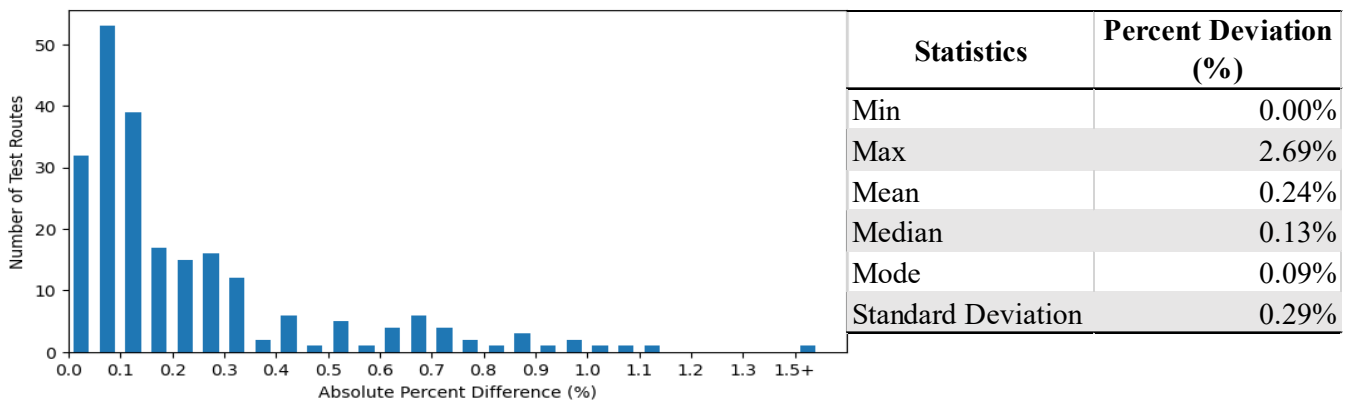

Figure 3. START population data comparison to calculated population data using latest LandScan data inputs

| Statistics | Percent Deviation (%) |
|---|---:|
| Min | 0.00% |
| Max | 2.69% |
| Mean | 0.24% |
| Median | 0.13% |
| Mode | 0.09% |
| Standard Deviation | 0.29% |

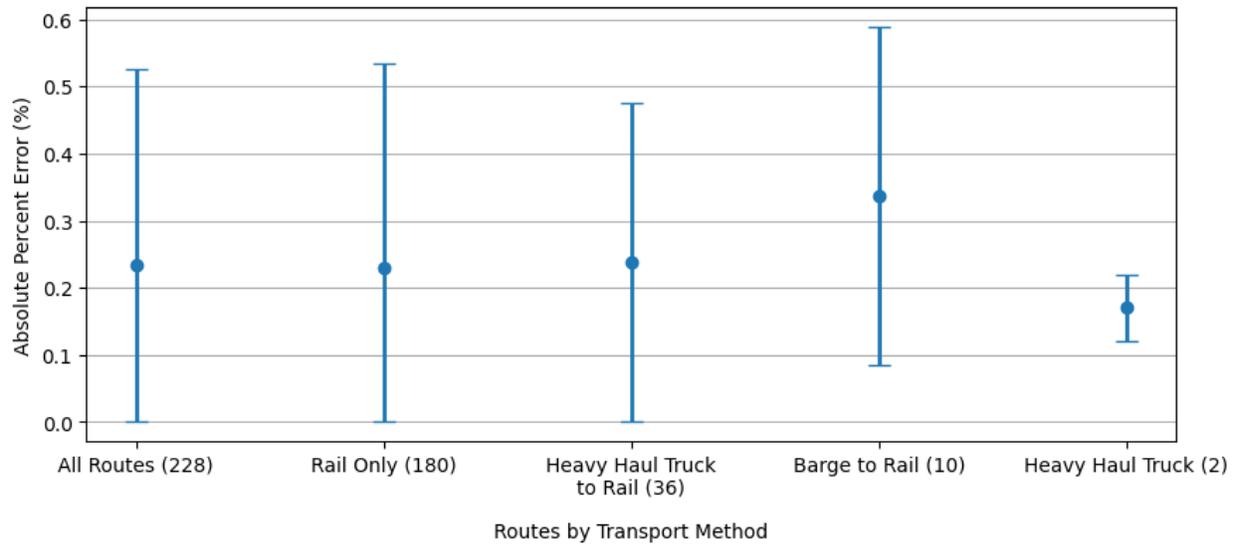

**Figure 4. START population data comparison by transport method**

## START ROUTE LENGTH DATA

Route length was identified as a crucial output of START that would be used in an impact evaluation, so it is a feature that is checked by the V&V efforts. The route distance was verified and validated using the QGIS software using the KML file of the route downloaded from START. In order to determine the route length in QGIS, the route must be reprojected from the native ESPG:4326 projection into ESPG:3857 or web Mercator in order to calculate the route distance (the population calculations are all conducted in the native raster projection of ESPG:4326) which is why the downloaded KML route must first be reprojected. This can be accomplished in QGIS by applying the projection tool to convert the polyline route from the native projection to a user-defined projection. The user can then go into the attribute table of the reprojected route and use the field calculator tool to obtain the route distance. For this evaluation, automated python scripts were developed in QGIS and used to calculate the route length.

The route distance V&V was conducted in QGIS because it is fully independent of START. In addition, START utilizes the ArcMap Network Analyst tool for this calculation. The route distance calculations available to the user in START are not easily replicated in the user interface of ArcMap without the Network Analyst tool. The user would need to develop a python work-around to replicate this calculation in ArcMap in the same projection that START utilizes.

Figure 5 shows the START route length data comparison as calculated in QGIS as well as the statics associated with that analysis. Figure 6 shows START length data comparison by transportation method to visualize if any one transportation mode The results showed overall good agreement with an average difference of 0.53% (SD 0.45%). Routes were then separated into subsets by transportation method to determine if there were any correlations. All subsets had similar differences.

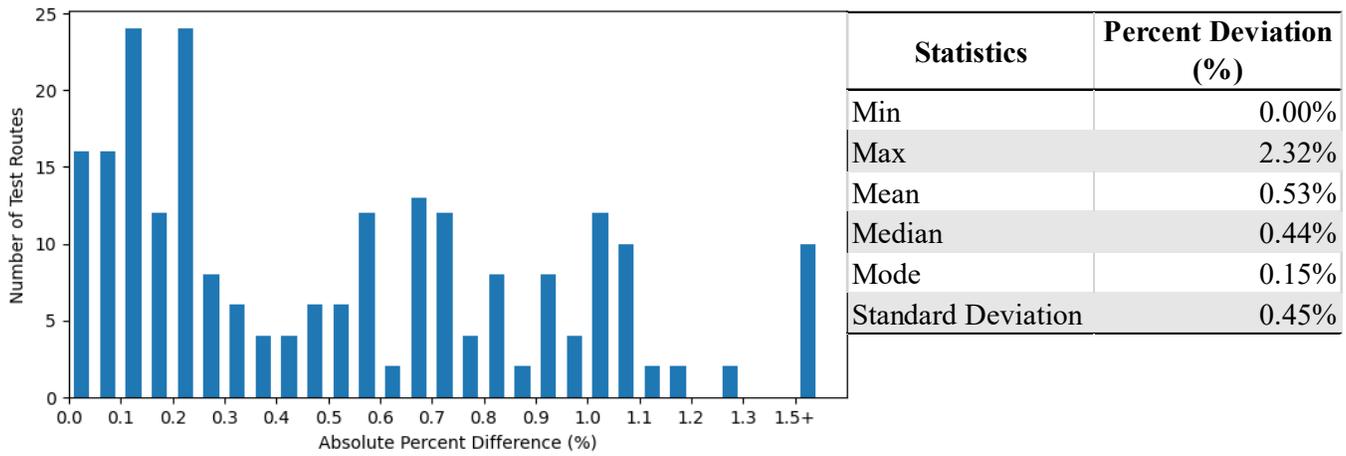

**Figure 5. START length data comparison as calculated in QGIS**

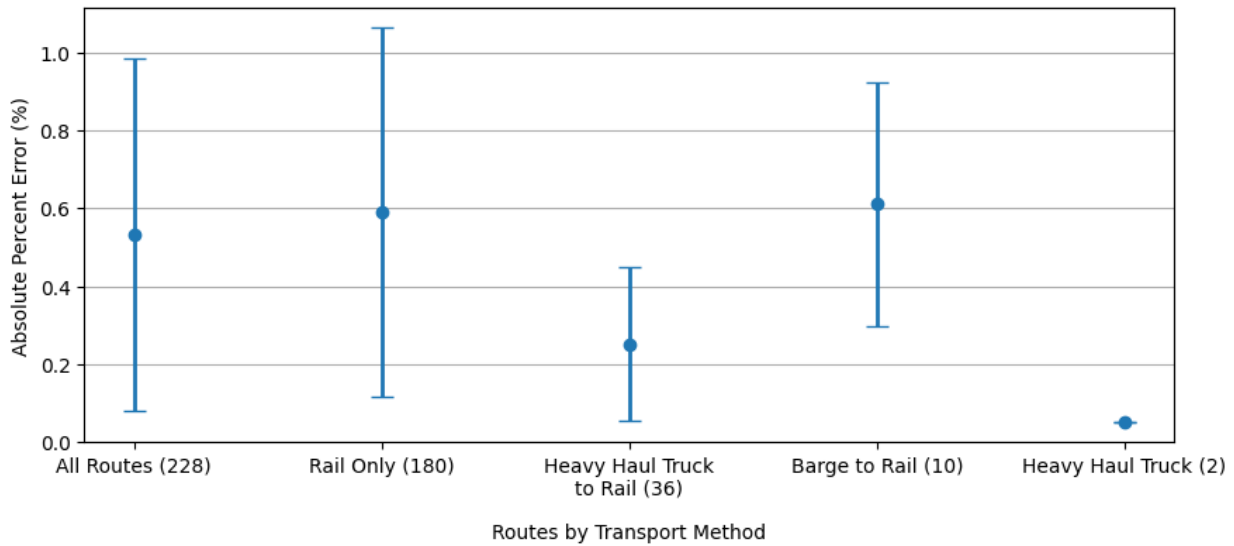

**Figure 6. START length data comparison by transportation method**

**CONCLUSIONS**

As DOE continues to plan for the eventual transportation, storage, and disposal of SNF and HLW from nuclear power plants and DOE sites, tools assisting in decision making and risk assessment will be needed. START provides a tool that provides stakeholders with the ability to compare routes attributes such as accident likelihood, expected doses to population, or length of route among others. With each update, START data is verified by an independent V&V team.

With the update of START to v.3.2.2, the V&V team verified length and population using independent data. The population data was verified using the latest LandScan data in QGIS and show good agreement with an average difference of 0.24% (SD 0.29%). The length data was compared to data generated in QGIS and showed good agreement with an average difference of 0.53% (SD 0.45%). Future V&V efforts will focus on continued verification of length and population data as well as validation of population density and collective dose to populations adjacent to the route.

**DISCLAIMER**

This is a technical report that does not take into account contractual limitations or obligations under the Standard Contract for Disposal of Spent Nuclear Fuel and/or High-Level Radioactive Waste (Standard Contract) (10 CFR Part 961).

To the extent discussions or recommendations in this report conflict with the provisions of the Standard Contract, the Standard Contract governs the obligations of the parties, and this report in no manner supersedes, overrides, or amends the Standard Contract.

This report reflects technical work which could support future decision making by DOE. No inferences should be drawn from this report regarding future actions by DOE, which are limited both by the terms of the Standard Contract and Congressional appropriations for the Department to fulfill its obligations under the Nuclear Waste Policy Act including licensing and construction of a spent nuclear fuel repository.